\documentclass[%
 reprint,
 amsmath,amssymb,
 aps,
]{revtex4-2}

\usepackage{graphicx}
\usepackage{dcolumn}
\usepackage{bm}
\usepackage{xcolor}

\begin{document}

\preprint{APS/123-QED}


\title{Misinference of interaction-free measurement from a classical system}

\author{Valeri Frumkin}
 \email{valerafr@mit.edu}
\affiliation{Department of Mathematics, Massachusetts Institute of Technology.}

\author{John W. M. Bush}
 \email{bush@math.mit.edu}
\affiliation{Department of Mathematics, Massachusetts Institute of Technology.}


\begin{abstract}
Interaction-free measurement is thought to allow for quantum particles to detect objects along paths they never traveled. As such, it represents one of the most beguiling of quantum phenomena. Here, we present a classical analog of interaction-free measurement using the hydrodynamic pilot-wave system, in which a droplet self-propels across a vibrating fluid surface, guided by a wave of its own making. We argue that existing rationalizations of interaction-free quantum measurement in terms of particles being guided by wave forms allow for a classical description manifest in our hydrodynamic system, wherein the measurement is decidedly not interaction-free.
\end{abstract}

\maketitle

Interaction-free measurement in quantum mechanics seemingly allows one to obtain information about the quantum state of an object without its being “disturbed” by the measurement process. Similar ideas, such as negative result measurement, can be traced back to Renninger \cite{renninger_messungen_1960} and Dicke \cite{dicke_interactionfree_1981}. However, the currently accepted notion of interaction-free measurement is due to Elitzur and Vaidman \cite{elitzur_quantum_1993}, who argued that the presence of an object in an interferometer could modify the interference of a photon, even without any direct interaction between object and photon. 
The authors considered a Mach-Zehnder interferometer with a bomb placed along one of its arms (figure 1). The bomb has a highly sensitive trigger, so that any photon passing through it will detonate the bomb; however, if the bomb malfunctions, the photon passes undisturbed. Since a functioning or malfunctioning bomb is equivalent to its being present or absent in the interferometer, we adopt the latter description.
The standard rationale for the effect is as follows \cite{elitzur_quantum_1993}. As the photon enters the first beam-splitter $B_1$, its wave function is split in two, with each part traveling along one arm of the interferometer. In the absence of a bomb, the two wave packets recombine at the second beam-splitter $B_2$, in which case the photon will be detected at detector $D_1$ with probability $1$. If the bomb is present, $50\%$ of the time the particle follows path 1 and detonates the bomb; otherwise, it follows path 2. As it does so, path 1 is blocked by the bomb, thus; interference between the two wave packets at $B_2$ is prevented, and the particle is detected at $D_1$ or $D_2$ with equal probability. Thus, if a particle is detected at $D_2$, it provides the experimenter information about the presence of a live bomb positioned along a path that it never traveled. After many realizations of the experiment in which the bomb is present $50\%$ of the time, the experimenter has a $25\%$ chance of detecting the particle at $D_2$ and so the bomb along path 1.

\begin{figure} [b]
\noindent \begin{centering}
\hspace*{-0.5cm}
\includegraphics[width=20pc]{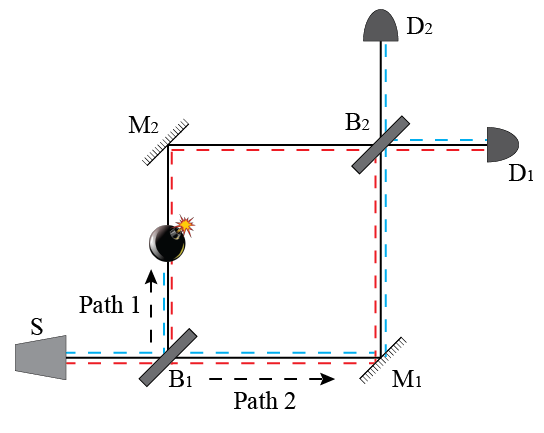}
\par\end{centering}
\caption{A schematic of the Elitzur-Vaidman bomb experiment. A particle  emitted from a source $S$ passes through a beam-splitter $B_1$, at which point its associated wave (a wave function \cite{elitzur_quantum_1993} or a pilot-wave \cite{hardy_existence_1992}) is split in two. The wave is then recombined at a beam-splitter $B_2$, and the particle continues toward the detectors. In the absence of a bomb, the particle will be detected at $D_1$ $100\%$ of the time, while in its presence, the particle will be detected at $D_2$ $25\%$ of the time. A detection event at $D_2$ indicates the presence of a live bomb along a path the particle never took. Red and blue dashed lines indicate possible paths taken by a particle emitted from $S$ in the absence and presence of the bomb, respectively.}
\label{fig:1}
\end{figure}

\begin{figure*} [t!]
\includegraphics[width=42pc]{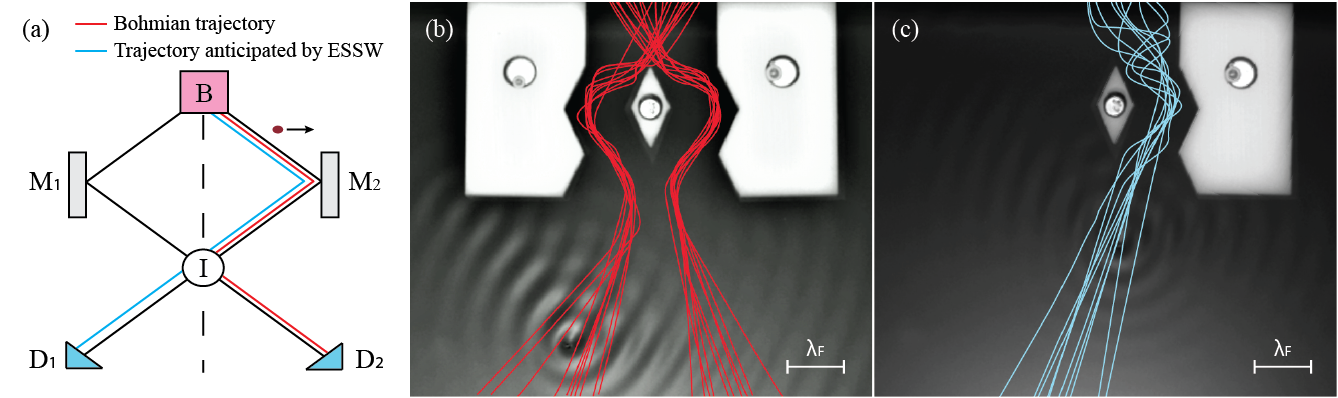}
\caption{Surreal trajectories in quantum mechanics and pilot-wave hydrodynamics~\cite{frumkin_real_2022}: (a) A variant of the interferometer setup considered by Englert, Scully, S\"ussman, and Walther (ESSW) \cite{Englert1993}. An incoming wave packet is split by a beam splitter $B$ and reflected by the mirrors $M_1$ and $M_2$. The wave packets interfere in the region $I$ and then move towards the detectors $D_1$ and $D_2$. The blue path represents the particle trajectory anticipated by ESSW, while the red path is that predicted by Bohmian mechanics, the so-called ``surreal trajectory". The red dot represents the one-bit, which-way detector employed in the weak-measurement experiments of Mahler {\it et al.~\cite{mahler_experimental_2016}}.
(b) In the associated hydrodynamic analog~\cite{frumkin_real_2022}, droplets reflect off submerged barriers, indicated in white. When the setup is symmetric, the droplet enters the right or left channel with equal probability and is then reflected off the associated barrier. Its subsequent deflection away from the system centerline results in a real surreal trajectory. 
Twenty such trajectories are shown. (c) When one of the barriers is removed, the symmetry of the system is broken. The walking droplet is then reflected away from the remaining barrier, resulting in the trajectory that one might expect. }
\label{fig:2}
\end{figure*}

Interaction-free measurement has been studied in various contexts, both theoretically \cite{mitchison_absorption-free_2001, ryff_null-result_2014,zhou_nondistortion_2001,simon_fundamental_2000,elitzur_nonlocal_2001,hardy_existence_1992, karlsson_interaction_1998,kwiat_interaction-free_1995,thomas_semitransparency_2014,van_voorthuysen_quantum-mechanical_1997}, and experimentally \cite{kwiat_experimental_1995,du_marchie_van_voorthuysen_realization_1996,hafner_experiment_1997,tsegaye_efficient_1998,white_interaction-free_1998,kwiat_high-efficiency_1999,elitzur_nonlocal_2001,inoue_experimental_2000,turner_interaction-free_2021,peise_interaction-free_2015}. However, there is still no widely accepted consensus on the precise meaning of the term ``interaction-free" \cite{vaidman_meaning_2003}. Indeed, the extent to which an experiment like that proposed by Elitzur and Vaidman is interaction-free depends on the particular interpretation of quantum mechanics one chooses to adopt. 
Notably, the Copenhagen interpretation of quantum mechanics provides no description of particles traveling along well defined paths, so is at odds with the standard description of the effect, which relies explicitly on the notion of particle trajectories. 
An interpretation that does allow one to describe real particle paths is Bohmian Mechanics, also known as the de Broglie-Bohm pilot-wave theory \cite{bohm_undivided_1993,holland_quantum_1993,durr_bohmian_2009}. According to Bohmian mechanics, particles move under the influence of the wave function, with a velocity equal to the quantum velocity of probability in the standard quantum formalism. 
Both the particle and its pilot-wave are assumed to be real physical objects. The Bohmian version of the Elitzur-Vaidman experiment, was first presented by Hardy \cite{hardy_existence_1992}, who posited the following physical picture. At the first beam splitter, the particle takes one of the two paths in the interferometer, while its pilot-wave is split in two. One part of the pilot-wave carries the particle along its path, while the other, so-called {\it empty wave}, proceeds along the other path without the particle. Both components of the pilot-wave recombine at $B_2$, producing the standard outcome. If a bomb is present along the path that the particle didn't take, it blocks the empty wave, preventing it from interfering with the particle's pilot-wave at $B_2$, thus enabling the particle to arrive at detector $D_2$ with non-zero probability. 
Note that according to both this physical picture and that proposed by Elitzur and Vaidman \cite{elitzur_quantum_1993}, the experiment can be considered ``interaction-free" only in the sense that the particle does not interact with the bomb: the wave form propagating along Path 1 interacts with and is altered by the bomb without detonating it. We proceed by demonstrating that such detection of an object by a particle that interacts with the particle only through its associated waveform may likewise be achieved in a classical system. We then argue that the term ``interaction-free" is misleading, both for the classical system under consideration and its quantum counterpart.

A classical pilot-wave system was discovered by Yves Couder in 2005~\cite{couder_walking_2005}, and consists of a millimetric droplet bouncing and self-propelling across the surface of a vibrating liquid bath through an interaction with its own wave. 
This system has provided the basis for a growing list of hydrodynamic quantum analogs \cite{bush_pilot-wave_2015,bush_hydrodynamic_2020}, including single- and double-slit diffraction and single-particle interference \citep{couder_single-particle_2006}, quantization of orbital states \cite{fort_path-memory_2010,perrard_self-organization_2014}, the emergence of wave-like statistics in corrals \cite{harris_wavelike_2013, saenz_statistical_2018}, Friedel oscillations \cite{saenz_hydrodynamic_2020}, superradiance \cite{frumkin_superradiant_2023}, and hydrodynamic spin lattices \cite{saenz_emergent_2021}.  
As the bouncing droplets are the sources of their own pilot-wave field, the accompanying physical picture is more closely aligned with the relativistic double-solution pilot-wave theory proposed in the 1920s by 
Louis de Broglie~\cite{deBroglie1987} than with the de Broglie-Bohm theory~\cite{bush_hydrodynamic_2020}. Nevertheless, the hydrodynamic pilot-wave system has captured certain features predicted by Bohmian mechanics, including surreal particle trajectories \cite{frumkin_real_2022}, 
unexpected trajectories predicted by Bohmian mechanics to arise in an interferometer owing to the interference of empty waves with the particle's pilot wave (see Figure 2). While surreal trajectories were 
thought to be unphysical~\cite{Englert1993}, Mahler {\it et al.~\cite{mahler_experimental_2016} reported that experimental trajectories inferred from weak-measurement are consistent with the predictions of Bohmian mechanics.} 
As noted by Vaidman~\cite{vaidman_reality_2005}, the phenomenon of surreal trajectories is related to that of interaction-free measurement, in that both rely on interference effects between a particle-carrying pilot-wave and an empty wave.
Here, we build upon the experimental setup used to realize surreal trajectories in pilot-wave hydrodynamics in order to demonstrate that when a particle is accompanied by a pilot-wave, an analog of interaction-free measurement can be realized in a classical system.

\begin{figure*} [t!]
\includegraphics[width=42pc]{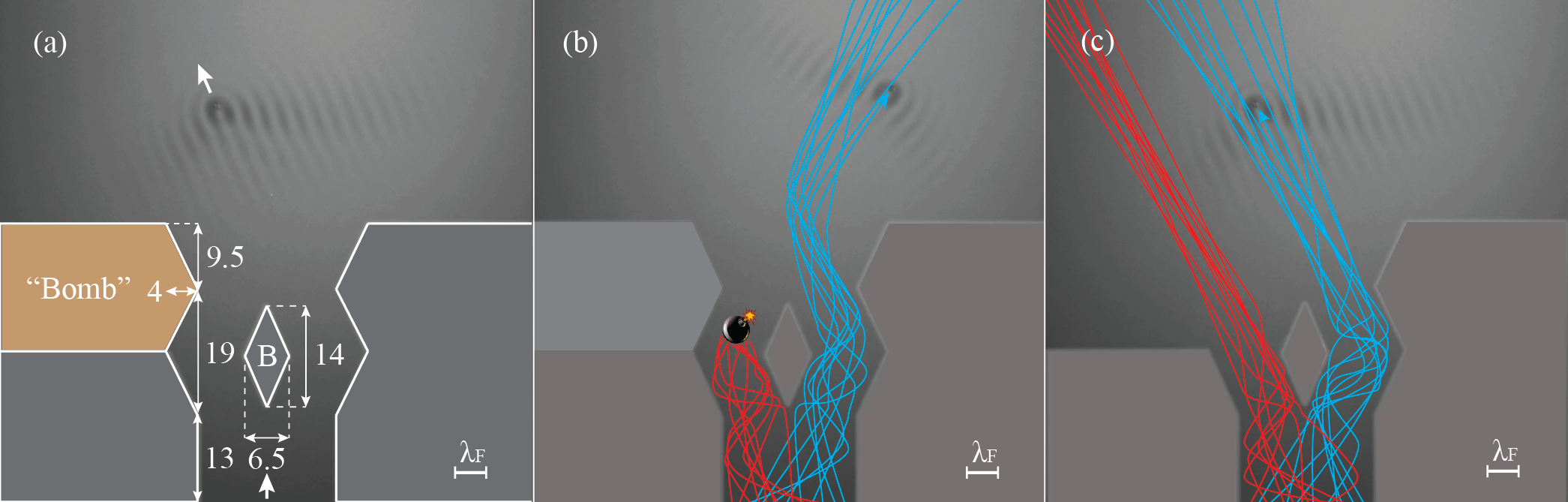}
\caption{Interaction-free measurement in pilot-wave hydrodynamics: (a) A schematic illustration of the topography used in the experiment. The upper half of the left barrier (orange) plays the role of a ``bomb".
(b) In a symmetric setup, the droplet enters the right or left channel with equal probability.  If the droplet goes to the left channel, it detonates the ``bomb", and if it goes to the right, it is deflected away from the system centerline, resulting in a “surreal” trajectory \cite{frumkin_real_2022}. Thus, the droplet will always be detected on the right side of the setup. Twenty such trajectories are shown. (c) If the ``bomb" is removed, the symmetry of the system is broken, the droplet's pilot-wave does not interact with the bomb, and ``surreal" trajectories are suppressed. Thus, the droplet will always be detected on the left side of the setup. If the bomb is present $50\%$ of the time, there is $25\%$ chance of the droplet being detected on the right, and so the bomb on the left. The scale bar represents the Faraday wavelength $\lambda_{F}=4.84$ mm }
\label{fig:3}
\end{figure*}

Our experimental system consists of a circular bath filled with a $7.0\pm 0.3$ mm deep layer of silicon oil with surface tension $\sigma=0.0209$ N/m, viscosity $20$ cSt, and density $\rho=0.965\times 10^{-3}\,\,\,\text{kg}/\text{m}^{3}$. 
The height of the submerged topography is $6.0\pm 0.2$ mm, so the oil depth in the overlaying shallow regions is approximately $1$ mm. 
The bath is vibrated vertically using an electromagnetic shaker with forcing $F(t)=\gamma\cos(2\pi ft)$, peak vibrational acceleration $\gamma=3.95$g and frequency $f=80$ Hz. The Faraday threshold, at which the  free surface destabilizes to a pattern of subharmonic Faraday waves, was $\gamma_F=4.05$g. 
The electromagnetic shaker was connected to the bath by a steel rod coupled with a linear air bearing, thus insuring spatially uniform vibration \cite{harris_generating_2015}. Two accelerometers placed on opposite sides of the bath allowed for control of the vibrational acceleration amplitude to within $\pm 0.002$ g.
We visualized the droplet trajectories and guiding pilot-wave using a semi-reflective mirror angled at 45° relative to the bath, and a charge-coupled device (CCD) camera mounted directly above the setup. We used a diffuse-light lamp, positioned horizontally in front of the mirror, to illuminate the bath, yielding images with bright regions corresponding to horizontal regions of the bath surface, specifically extrema or saddle points.

The topographical configuration used in our experiments is depicted in figure 3(a). A walking droplet is launched towards a submerged rhombus that acts as a beam-splitter, forcing the droplet along one of two adjoining channels with equal probability. We consider the upper part of the left channel to be our ``bomb", an object whose presence we seek to detect without the droplet interacting with it directly.
While the droplet is localized at all times,  its pilot-wave is spatially extended and so interacts with the geometry of its environment. This delocalization of the pilot-wave accounts for the different drop behaviors in the different geometric configurations.
When the ``bomb" is present in the system (figure 3(b)), the setup is identical to that used to realize surreal trajectories \cite{frumkin_real_2022}. If the droplet takes the left channel, it hits the ``bomb" and so effectively detonates it.  At this point, the droplet trajectory is considered to be terminated.
If, however, the droplet takes the right channel, its pilot-wave interacts with its environment so as to deflect the droplet toward the right, along a surreal trajectory. Thus, if the ``bomb" is present, the droplet will either detonate it, or be detected on the right side of the setup (see supplementary video 1).  The droplet follows a surreal trajectory followed only when the boundary geometry of the setup is right-left symmetric \cite{frumkin_real_2022}.
When the ``bomb" is absent (Figure 3(c)), this symmetry is broken, and the surreal trajectory is precluded.
If the droplet takes the right path, it is reflected, then continues in a straight line towards the left. If the droplet takes the left path, there is nothing there to reflect it, so it continues along a straight path to the left.
Thus in the absence of the ``bomb", the droplet will always be detected on the left side of the setup (see supplementary video 2). After many realizations of the experiment in which the bomb was present $50\%$ of the time, the experimenter has a $25\%$ chance of detecting the droplet on the right side. Such a detection indicates that the ``bomb" was present in the left channel, even though the droplet took the right path and so never interacted directly with it. 

{\it Discussion:} Our experiment demonstrates that if particles are accompanied by guiding wave forms, the statistical behaviour that has led to the inference of interaction-free measurement in quantum mechanics may be achieved in a classical system.
Notably, not all interpretations of quantum mechanics need appeal to interaction-free measurement. In their original paper, Elitzur and Vaidman write that according to the many-worlds interpretation \cite{everett_relative_1957} ``since all worlds take place in the physical universe, we cannot say that nothing has touched the object. We get information about the object without touching it in one world, but we pay the price of interacting with the object in the other world" \cite{elitzur_quantum_1993}. 
In Bohmian mechanics, while the particle does not interact with the bomb, its pilot-wave is blocked by it, so the measurement is not strictly interaction-free.
In the context of the Copenhagen interpretation of quantum mechanics, before measurement, one can only talk about the particle's wave function, which is a nonlocal object that is determined by the configuration of the interferometer as a whole. Thus, in the absence of a bomb, the wave function can only collapse at detector $D_1$, while in its presence, the wave function has a non-zero probability of collapsing at $D_2$. According to this interpretation, the wave function is considered to be a mathematical construction for computing probabilities rather than a physical field; thus,  one could argue that the effect is indeed interaction-free. However, a similar argument can be 
applied quite generally in the Copenhagen picture, according to which the measurement process is marked by the collapse of the wave function, and the probability of a particular outcome is prescribed by the configuration of the system as a whole. 
For example, according to the Copenhagen interpretation, any measurement of one half of an entangled pair will yield information about the other half without interacting with it. Thus, all such measurement outcomes may be considered interaction-free, rendering the notion superfluous. If, however, one thinks of particles in the Elitzur-Vaidman experiment as physically localized wave-packets traveling in an interferometer, a more accurate description of the effect would be ``measurement-free interaction", since the presence of the bomb is presumed not to collapse the wave function.

It is important to note the differences between interaction-free measurement in quantum mechanics, and the analogous statistical inference made in the classical system presented here. First, the quantum wave function is a nonlocal object that is determined by the entire configuration of the experimental setup. As a result, interaction-free measurement can be realized (at least in theory) with an interferometer that has arms of unequal length, provided the lengths of the two arms ($B_1M_1B_2$ and $B_1M_2B_2$ in Figure 1) differ by $n\lambda$, and that this length difference is smaller than the coherence length of the interferometer.
In our system, the pilot-wave is affected by the global geometry, just as a standing field of Faraday waves is affected by boundaries  
in confined geometries
\cite{kudrolli_scarred_2001}. The form of the pilot wave is affected by the totality of the boundary geometry only if its spatial extent is sufficiently large.
Consequently, changing the configuration of the setup (for example, by either increasing its size or decreasing the system memory) will serve to suppress the surreal trajectories and so nullify the effect.  Second, in the Bohmian version of the Elitzur-Vaidman experiment, the bomb blocks the empty wave, thus eliminating interference.
In our setup, the situation is reversed: the presence of the ``bomb" promotes interference of the pilot-wave, resulting in a surreal trajectory. This configurational difference does not alter the key statistical behavior common to the two systems. Third and most importantly, whatever the case may be in its quantum counterpart, the statistical inference made in our system should in no way be taken as evidence of interaction-free measurement.

All current attempts to rationalize interaction-free measurement rely on a physical picture where localized wave-particle objects (e.g. traveling wave packets, or Bohmian particles with their pilot and empty waves) travel along the arms of the interferometer. 
Our study demonstrates that such localized wave-particle descriptions of interaction-free measurement can also be achieved in a classical system. We note that the validity of such descriptions can be tested directly along the lines first suggested by Renninger \cite{renninger_zum_1953}, by realizing the Elitzur-Vaidman bomb experiment while increasing the length of one of the interferometer arms by an amount $n\lambda$, where $n$ is an integer and $\lambda$ the wavelength of the photon. 
If one represents the photon as a wave packet that is split in two at $B_1$, then for sufficiently large $n$, the two wave packets will not recombine at $B_2$ since they travel paths of different length at the same speed. 
Likewise, in the Bohmian picture suggested by Hardy \cite{hardy_existence_1992}, the interference of the particle's pilot wave with an empty wave at $B_2$ would be eliminated.
Thus, the photon would have a non-zero probability of being detected at $D_2$ even in the absence of a bomb.
If the effect does not persist for arbitrary large $n$, it would validate these localized wave-particle descriptions and, when considered in light of our results, suggest that interaction-free measurement is a misnomer.
If the effect does persist for arbitrary large $n$ (as one might expect), this experiment would invalidate the localized wave-packet description of the problem. 
One would then be obliged to adopt a nonlocal view,
for example that offered up by the Copenhagen Interpretation or by the second-order formulation of Bohmian mechanics, according to which particles are guided by a quantum potential prescribed by the wave function
\cite{bohm_undivided_1993,holland_quantum_1993,durr_bohmian_2009}). Since the wave function is a nonlocal object, so too is the quantum potential; thus, the path taken by a Bohmian particle is determined by the entire configuration of the interferometer. 

As was the case in the hydrodynamic analog of surreal trajectories, the hydrodynamic pilot-wave system captures a feature of quantum mechanics correctly predicted by Bohmian mechanics, without having to invoke 
the quantum potential. The role of the quantum potential in guiding the particle in Bohmian mechanics is played by the droplet's guiding wave, which is generated by the droplet but interacts with the system geometry in such a way as to produce the surreal trajectories. In neither Bohmian mechanics nor our system is the measurement really interaction-free. In Bohmian mechanics, the quantum potential changes form according to the presence or absence of the bomb; in our system, the local, 
particle-generated pilot-wave does likewise.

Finally, given the numerous applications proposed for interaction-free measurement \cite{karlsson_interaction_1998,guo_quantum_1999,czachor_quantum_1999,mitchison_counterfactual_2001}, it is noteworthy that none of these requires the non-local features of the wave function for their implementation, only its interference properties. It is thus worth considering whether some of these applications might be implemented in the hydrodynamic pilot-wave system. A classical macroscopic implementation of these ideas could potentially allow one to bypass the difficulty of decoherence inherent in all quantum systems, and so conceivably serve as a physical platform for quantum-inspired classical computing.

\bibliographystyle{ieeetr}
\bibliography{InteractionFree.bib}

\vspace{0.5cm}
\subsection*{Acknowledgments} 
\subsubsection*{Funding}
The authors gratefully acknowledge the financial support of the National Science Foundation through grant CMMI-2154151.

\subsubsection*{Competing interests}
The authors declare no competing interests.

\subsubsection*{Data and materials availability}
All data are available in the main text or the supplementary materials.

\end{document}